\documentstyle[11pt,epsfig]{article}
\newcommand{\zp}[3]{Z. Phys.\ C#1 (19#2) #3}
\newcommand{\pl}[3]{Phys.\ Lett.\ B#1 (19#2) #3}
\newcommand{\np}[3]{Nucl.\ Phys.\ B#1 (19#2) #3}
\newcommand{\prd}[3]{Phys.\ Rev.\ D#1 (19#2) #3}
\newcommand{\prl}[3]{Phys.\ Rev.\ Lett.\ #1 (19#2) #3}

\def\simgt{\rlap{\lower 3.5 pt \hbox{$\mathchar \sim$}} \raise 1pt \hbox {$>$}}
\def\simlt{\rlap{\lower 3.5 pt \hbox{$\mathchar \sim$}} \raise 1pt \hbox {$<$}}

\newcommand{\beq}{\begin{equation}}
\newcommand{\eeq}{\end{equation}}
\newcommand{\bea}{\begin{eqnarray}}
\newcommand{\eea}{\end{eqnarray}}

\catcode`\@=11

\def\section{\@startsection{section}{1}{\z@}{3.5ex plus 1ex minus .2ex}
{2.3ex plus .2ex}{\large\bf}}

\def\thesection{\arabic{section}.}

\def\appendix{\setcounter{section}{0}
 \def\thesection{Appendix \Alph{section}:}
 \def\theequation{\Alph{section}.\arabic{equation}}}

\def\@citex[#1]#2{\if@filesw\immediate\write\@auxout{\string\citation{#2}}\fi
  \def\@citea{}\@cite{\@for\@citeb:=#2\do
    {\@citea\def\@citea{,\penalty\@m}\@ifundefined
       {b@\@citeb}{{\bf ?}\@warning
       {Citation `\@citeb' on page \thepage \space undefined}}%
\hbox{\csname b@\@citeb\endcsname}}}{#1}}

\def\citer{\@ifnextchar [{\@tempswatrue\@citexr}{\@tempswafalse\@citexr[]}}

\def\@citexr[#1]#2{\if@filesw\immediate\write\@auxout{\string\citation{#2}}\fi
  \def\@citea{}\@cite{\@for\@citeb:=#2\do
    {\@citea\def\@citea{--\penalty\@m}\@ifundefined
       {b@\@citeb}{{\bf ?}\@warning
       {Citation `\@citeb' on page \thepage \space undefined}}%
\hbox{\csname b@\@citeb\endcsname}}}{#1}}
\relax
\begin{document}

\thispagestyle{empty}

\hfill\vbox{\hbox{\bf DESY 95-098}
            \hbox{\bf Fermilab-Pub-95/131-T}
                                }

\vspace{0.5in}
\begin{center}
\boldmath
{\Large\bf Cross Sections for Charm Production in $ep$ Collisions:\\[2mm]
           Massive versus Massless Scheme} \\
\unboldmath
\vspace{0.5in}

{\large \sc B.A.~Kniehl$^{a,*}$, M.~Kr\"amer$^{b}$, G.~Kramer$^{c}$,
            M.~Spira$^{c}$}

\vspace{0.5in}

$^a$Fermi National Accelerator Laboratory, Batavia, IL 60510, USA\\[2mm]
$^b$Deutsches Elektronen-Synchrotron DESY, D-22603 Hamburg, Germany\\[2mm]
$^c$II. Institut f\"ur theoretische Physik$^{\dagger}$, Universit\"at
    Hamburg, D-22761 Hamburg, Germany

\end{center}

\vspace{0.5in}

\begin{abstract}
The next--to--leading order inclusive cross section for large-$p_\perp$
photoproduction of charm quarks at HERA is calculated in two different
approaches. In the first approach the charm quarks are treated as massive
objects which are strictly external to the proton and the photon while
in the second approach the charm mass is neglected and the $c$ quark is
assumed to be one of the active flavours in the proton and photon structure
functions. We present single-inclusive distributions in transverse momentum
and rapidity including direct and resolved photons. The cross section in
the massless approach is found to be significantly larger than in the massive
scheme. The deviation originates from several contributions which are
disentangled. We argue that large-$p_\perp$ photoproduction of charm quarks
at HERA will be sensitive to the charm content of the photon structure
function.
\end{abstract}

\vfill

\noindent
{\footnotesize
$\dagger\,$ Supported by Bundesministerium f\"ur Forschung und Technologie,
            Bonn, Germany, Contract 05 6 HH 93P (5) and EEC Program
            ``Human Capital and Mobility'' through Network ``Physics at High
            Energy Colliders'' under Contract CHRX-CT93-0357 (DG12 COMA).\\
$*\,$ Permanent address: Max-Planck-Institut f\"ur Physik,
      F\"ohringer Ring 6, 80805 Munich, Germany.}

\newpage

\section{Introduction}
Charm production in high energy $ep$ collisions at HERA is dominated by
photoproduction events where the electron is scattered by a small angle
producing photons of almost zero virtuality ($Q^2 \simeq 0$). In leading
order (LO) QCD the main process is photon-gluon fusion where the photon
interacts directly with the gluon from the proton producing a $c\bar{c}$
pair in the final state ($\gamma + g \to c + \bar{c}$). Besides the direct
photoproduction channel charm production at HERA can proceed also via the
resolved photoproduction process where the photon behaves as a source of
partons which interact with the partons in the proton, as for example in
$g + g \to c + \bar{c}$. It is known that for the photoproduction of light
quarks $q$ (i.e. $q=u,d$ and $s$) and gluons the cross section of the
resolved process is much larger than the cross section for direct
photoproduction \citer{drees,ks}. Only at rather large transverse momenta,
$p_\perp\; \simgt\; 30$~GeV, the cross sections for the two types of
processes are comparable. This has been confirmed in various ways by
measurements of the ZEUS \cite{zeus} and H1 \cite{h1} collaborations at HERA.
Actually a recent comparison of the ZEUS experimental data for the inclusive
one-jet cross section \cite{zeus1} with the next-to-leading order (NLO)
predictions superimposing resolved and direct photoproduction shows
satisfactory agreement in absolute normalization, $p_\perp$ and rapidity
dependences \cite{kks}. These predictions are meant to be valid only for
large enough $p_\perp$, i.e. $p_\perp\; \simgt\; 5$~GeV, so that soft
processes, which can not be calculated in perturbative QCD, are negligible.
In the calculations the authors assume four active flavours, $q = u,d,s$ and
$c$, which are taken to be massless. This means that also charm quarks are
produced and counted as jets. Of course, the $c$ quark is also an ingoing
parton originating either from the proton or the photon in the case of the
resolved contribution. To neglect the charm mass $m$ in these calculations
is a reasonable approximation since in large-$p_\perp$ jet production the
effective scale is $p_\perp$ which is much larger than $m \simeq 1.5$~GeV.
The same approach is followed for predicting inclusive cross sections for
the photoproduction of light hadrons, like $\pi$'s, $K$'s etc., when the
produced massless $c$ quarks fragment into the light hadrons similarly as
the produced $u,d,s$ quarks and the gluons \cite{bkk}. In all of these
theoretical investigations the $c$ quark is one of the massless active
flavours in the photon and proton structure functions. In the following we
shall refer to this as the massless charm scheme, which we assume to be valid
in the region of large transverse momenta $p_\perp \gg m$. In this scheme the
small $p_\perp$ region is not calculable. The cross section diverges in the
limit $p_\perp \to 0$ and total production rates can not be predicted.

In the massive charm scheme which has been adopted by many authors (for a
review see \cite{smith}) the charm mass, $m \gg \Lambda_{\mbox{\scriptsize
QCD}}$, acts as a cutoff and sets the scale for the perturbative
calculations. Similarly to the case of large-$p_\perp$ jet production
the cross section factorizes into a partonic hard scattering cross section
multiplied by light quark and gluon densities \cite{css}. Inherent in this
factorization is the notion that the only quarks in the hadron and the
photon are the light ones. Thus in the massive charm scheme the number of
active flavours in the initial state is equal to $n_f = 3$ and the massive
$c$ quark appears only in the final state. This approach has the advantage
that not only various distributions, like in rapidity and/or transverse
momentum, can be predicted but also the total cross section. In LO direct
production is described by the partonic reaction $\gamma + g \to c + \bar{c}$
while the resolved contribution involves the channels $q+\bar{q}\to
c+\bar{c}$ and $g+g\to c+\bar{c}$, where $q(\bar{q})$ are light (massless)
quarks. NLO corrections have been calculated to these processes and found to
be substantial \cite{ellis,dkzz}. A comparison between the theoretical NLO
results and experimental data on photoproduction of charm quarks in
low-energy $\gamma p$ and $\gamma\gamma$ collisions has shown reasonable
agreement \cite{rid}. Recently, the total charm production cross section at
HERA has been measured \cite{zeus2} and found to be well described by the
prediction of Frixione et al. \cite{frix}. Depending on the choice of the
photon structure function the resolved cross section seems to contribute
only a fraction to the total charm production cross section \cite{frix}.

One might expect that the massive approach is reasonable only in those
kinematical regions where the mass $m$ and any other energy scale like
$p_\perp$ are approximately of the same magnitude and significantly larger
than $\Lambda_{\mbox{\scriptsize QCD}}$. Under these circumstances the
charm mass is used as the scale of $\alpha_s$ and the quark and gluon
densities of the photon and the proton. In next-to-leading order terms $\sim
\alpha_s(\mu^2)\ln(p_\perp^2/m^2)$ arise from collinear emission of gluons by
a heavy quark at large transverse momentum or from almost collinear branching
of gluons or photons into heavy quark pairs. These terms are not expected to
affect the total production rates, but they might spoil the convergence of
the perturbation series and cause large scale dependences of the NLO result
at $p_\perp \gg m$. In the massive approach the prediction of differential
cross sections is thus limited to a rather small range of $p_\perp\sim m$.
The proper procedure for $p_\perp \gg m$ is to absorb the terms proportional
to $\ln(p_\perp^2/m^2)$ into the charm distribution functions of the incoming
photon and proton and into the fragmentation function of $c$ quarks into
charmed hadrons. Of course, to do this absorption one needs a charm
contribution in the structure function in the first place. An alternative
way of making predictions at large $p_\perp$ is to treat the charm quarks
as massless partons from the start. The mass singularities of the form
$\ln(p_\perp^2/m^2)$ are then absorbed into structure and fragmentation
functions in the same way as for the light $u$, $d$, $s$ quarks. We expect
this massless approach to be better suited for the calculation of the
differential $p_\perp$ distributions up to NLO in the region $p_\perp \gg
m$. Then the problem arises how to proceed in the intermediate range where
$p_\perp > m$.

In order to investigate the region where $p_\perp > m$ we have calculated
the differential cross section
$\mbox{d}^2\sigma/\mbox{d}y\,\mbox{d}p_\perp^2$
as a function of $p_\perp$ with the rapidity $y$ integrated over the region
$|y| \le 1.5$. We shall compare the results in the two approaches: (a) the
massive charm approach with $m = 1.5$~GeV, in which we have computed the
cross section for open charm production and (b) the massless approach, where
we have evaluated the same differential cross sections for inclusive charmed
particle production. In both calculations we include direct and resolved
processes and go up to NLO where necessary. The massive calculation is based
on the work presented in \cite{dkzz} while the massless predictions use the
results obtained in \cite{kk}. In the second approach we need the
fragmentation function of the $c$ quark into charmed hadrons. This is
approximated by $\delta{}(1-z)$ where $z = p_D/p_c$ is the scaled momentum of
the charmed hadron, meson or baryon. With this choice the LO results in the
massive scheme approach the LO massless results in the limit $m \to 0$ if we
restrict ourselves to the same parton subprocesses. They must differ,
however, in NLO where the limit $m \to 0$ is not possible due to the
unabsorbed mass singular $\ln(p_\perp^2/m^2)$ terms. Of course it is no
problem to incorporate more realistic $c$ fragmentation functions in both
schemes but the choice above should be sufficient to see the essential
differences of the two approaches. A similar study of the production of
large-$p_\perp$ hadrons containing bottom quarks in $p\bar{p}$ collisions
has been performed by Cacciari and Greco \cite{cacc}.

\newpage

\section{Comparison of Results}
We have calculated the cross section $\mbox{d}^2\sigma/\mbox{d}y\,\mbox{d}
p_\perp^2$ in the HERA laboratory frame with $E_p = 820$~GeV protons colliding
with $E_e = 26.7$~GeV electrons travelling in the ($+z$) direction ($y > 0$).
The virtual photon spectrum is described in the Weizs\"acker-Williams
approximation with the formula
\beq\label{wwa}
f_{\gamma/e}(x) = \frac{\alpha}{2\pi}\left[\frac{1+(1-x)^2}{x}
\ln\frac{Q_{\mbox{\scriptsize max}}^2}{Q_{\mbox{\scriptsize min}}^2}
+2m_e^2x\left(\frac{1}{Q_{\mbox{\scriptsize max}}^2}-
              \frac{1}{Q_{\mbox{\scriptsize min}}^2}\right)\right]
\eeq
with $Q_{\mbox{\scriptsize max}}^2 = 4$~GeV$^2$,
     $Q_{\mbox{\scriptsize min}}^2 = m_e^2x^2/(1-x)$
 and $x = E_\gamma/E_e$ in the interval $0.15 < x < 0.86$ as in \cite{zeus2}
which corresponds to $\gamma{}p$ c.m.\ energies of 115~GeV~$< W <$~275~GeV,
$m_e$ is the electron mass. Since the average $Q^2 \simeq 10^{-4}$~GeV$^2$ is
very small, the photons are essentially on-shell, so that the $Q^2$ dependence
of $\sigma_{\gamma{}p}$ and the longitudinal contribution in (\ref{wwa}) can
be neglected. For the proton structure function we use the MRS(G) set
\cite{mrs} which describes well the proton structure at small $x$ and is
adjusted in the intermediate $x$ range. In Fig.1a the results for
the cross section $\mbox{d}^2\sigma/\mbox{d}y\,\mbox{d}p_\perp^2$ averaged
over the rapidity range $|y| < 1.5$ are presented for the direct contribution
in the massless and in the massive scheme. Both LO and NLO predictions are
shown. In the LO cross section the same NLO parton structure function MRS(G)
has  been adopted, only the parton-parton scattering cross sections are
evaluated in LO. The two-loop formula for $\alpha_s$ is used with the
$\Lambda$ value taken from the MRS(G) fit:
$\Lambda^{(4)}_{\overline{\scriptsize \mbox{MS}}} = 255$~MeV.
Both NLO calculations have been performed in the $\overline{\mbox{MS}}$
factorization and renormalization schemes. The corresponding scales have been
set to $\mu = \sqrt{p_\perp^2 + m^2}$. It is clear that in the massive scheme
only three flavours are active in the initial state and in the evaluation
of $\alpha_s$ whereas in the massless scheme also the charm distribution in
the proton contributes to $c(\bar{c})$ production and $\alpha_s$ is
calculated using four active flavours. We observe that the predictions in
the massless and massive scheme are very similar. In LO there is some
difference, the LO massless cross section is approximately 20\% larger than
the massive cross section over the whole range of $p_\perp$ between 3 and
15~GeV. This small difference originates from the additional charm
contribution in the proton and the value of $\alpha_s$. In NLO the direct
contributions of the massless and massive theories yield almost identical
results for the $p_\perp$ distribution in the range $3 < p_\perp < 15$~GeV.
We have checked that this is true also for higher values of the transverse
momentum up to $p_\perp \sim 30$~GeV.

The situation is completely changed for the resolved part of the $c/\bar{c}$
production. This contribution has been calculated with the NLO photon
structure function of GRV \cite{grv} in the $\overline{\mbox{MS}}$ scheme.
The results for the $p_\perp$ distribution are plotted in Fig.1b.
We observe that the NLO corrections in the massless scheme are large and
increase the cross section by roughly 100\%. In the massive scheme we have
calculated only the LO result, which is between one and two orders of
magnitude smaller than the LO massless cross section in the $p_\perp$ range
between 3 and 15~GeV.\footnote{Similar results have been obtained in the LO
analysis presented by R.~Godbole at the Photon~95 conference, Sheffield, UK,
April 1995 \cite{drgod}.} NLO QCD corrections do not change this
order-of-magnitude suppression \cite{frix} and will therefore not be taken
into account for the resolved contribution in the massive scheme.
We have examined that the prediction for the massless resolved cross section
does not depend strongly on the particular choice of the NLO GRV
parametrization of the $c$ distribution in the photon. The NLO GRV
parametrization in the DIS$_\gamma$ scheme\footnote{To be consistent the
appropriate changes have been made in the NLO direct part from the
$\overline{\mbox{MS}}$ to the DIS$_\gamma$ scheme \cite{kk}.} yields similar
results as well as the parametrization by Aurenche et al. \cite{aurenche}.

We may ask where the strong contribution of the resolved cross section in the
massless scheme is hidden in the massive approach. It must be contained in
the $\ln(p_\perp^2/m^2)$ terms in the NLO direct cross section. In the NLO
massless direct cross section all these terms are absorbed either in the
structure function of the proton, the photon or the fragmentation function.
Therefore at NLO the definition of direct/resolved photoproduction is
ambiguous and only the sum of both contributions should be considered.
In Fig.1c we have thus plotted the sum of the NLO massive direct and the LO
massive resolved contribution from Figs.1a and b (denoted NLO/LO massive) and
the same sum for the massless theory taken also from Figs.1a and b
(called NLO/LO massless). Considering the same order in $\alpha_s$ these are
the two predictions in the massive and massless theories which should be
compared. We observe that the NLO/LO massless result is approximately 60\%
larger than the NLO/LO massive prediction. We have investigated this
difference in some detail. To begin with one has to remember that in the
massless theory the collinear singular contributions in the final charm
states are absorbed into the fragmentation function. This subtraction is not
performed in the massive theory. If we subtract the corresponding
$\ln(p_\perp^2/m^2)$ terms in the massive theory the NLO/LO massive
prediction is increased by 25\%, so that we are left still with a difference
of 35\%. This difference originates mainly from the charm content of the
photon which does not correspond exactly to the $\ln(p_\perp^2/m^2)$ terms
in the massive theory arising from the almost collinear branching of
photons into charm quark pairs. We have examined this further by substituting
the GRV charm photon structure function by the massive LO ``Bethe-Heitler''
approximation \cite{wit}. This change reduces the difference between the
massless and the massive NLO/LO predictions to a level of 10\%. Part of the
difference between the two approaches can thus be attributed to theoretical
uncertainties concerning the charm content of the photon. The remaining
$\ln(p_\perp^2/m^2)$ terms connected with the collinear singularities at
the proton leg are higher order in $\alpha_s$ and are definitely much less
important.

In Fig.1c we have plotted also the sum of the direct and the resolved
components for the massless theory including NLO corrections (denoted NLO
massless). The higher order corrections to the resolved process increase the
theoretical prediction by 35\% compared to the NLO/LO massless result.
Following the discussion above this contribution should correspond to a
large extent to the NNLO direct cross section in the massive scheme which
has not been calculated yet.

In Fig.2 the differential cross sections
$\mbox{d}^2\sigma/\mbox{d}y\,\mbox{d}p_\perp^2$ are shown for $p_\perp =
10$~GeV as a function of the rapidity $y$ for the direct (Fig.2a)
and the resolved contributions (Fig.2b). Comparing the results in the
massive and the massless approaches we observe the same pattern as for the
$p_\perp$ distribution, Figs.1a and b. From Fig.2a we can infer that both
the LO and NLO predictions for the direct contribution in the massive and
massless approaches are very close together. In Fig.2b, where the resolved
$y$ differential cross sections are plotted we see the very much reduced
cross section in the massive theory and the appreciable cross section in
the massless approach, where the NLO prediction is in general larger than the
LO result.

Fig.2c finally shows the sum of direct and resolved contributions in
the massless and massive schemes. The same relation between massive and
massless schemes is seen here as in the $p_\perp$ distribution in
Fig.1c.

\section{Conclusion}

We have observed that the cross section in the massless approach is
approximately 110\% larger than in the massive scheme. The main part
of that deviation can be attributed to the difference found already
in the comparison of the NLO/LO massless cross section with the NLO/LO
massive result. This difference originates from several sources
which have been disentangled. The remaining increase of the NLO massless
prediction is due to the NLO corrections to the resolved process. Such a
contribution corresponds mainly to a NNLO correction to the direct part
which has not been calculated yet. In this sense the massless theory includes
a significant part of the NNLO correction and should therefore yield a more
reliable prediction than the massive scheme. This conclusion is supported
by the observation that the massless prediction is much more stable under
scale variation than the massive result.

Compared to a similar study for $b$ production in $p\bar{p}$ processes
\cite{cacc} we find much larger deviations between the massive and the
massless approaches. This can be attributed to the much stronger influence
of charm in the photon as compared to the proton caused by the point-like
component of the photon structure function. Consequently a measurement of
large-$p_\perp$ charm production at HERA might provide information about the
charm content of the photon (cf.~\cite{drgod}). Such measurements will be
very instructive since theoretical opinions on that issue are rather divided
\cite{intrin}.

Our study was based on a scale-independent $\delta$-type fragmentation
function of the charm quarks. In the future we shall extend this by
using more realistic forms for the fragmentation including evolution
to higher scales. This latter effect in the fragmentation might compensate
part of the $\ln(p_\perp^2/m^2)$ absorption in the massless theory and
might thus bring the massless prediction nearer to the massive result.

\noindent
{\em Note added:} After completion of our work we have been informed about
a similar analysis by Cacciari and Greco \cite{cacc2}.

\vspace{3mm}

{\bf Acknowledgements.} We thank the authors of \cite{cacc2} for making
available to us their results prior to publication. One of us (BAK) is
indebted to the FNAL Theory Group for inviting him as a Guest Scientist
and for the great hospitality extended to him.



\newpage

\begin{figure}[htbp]

\vspace*{-0.5cm}
\hspace*{0.75cm}
\epsfig{%
file=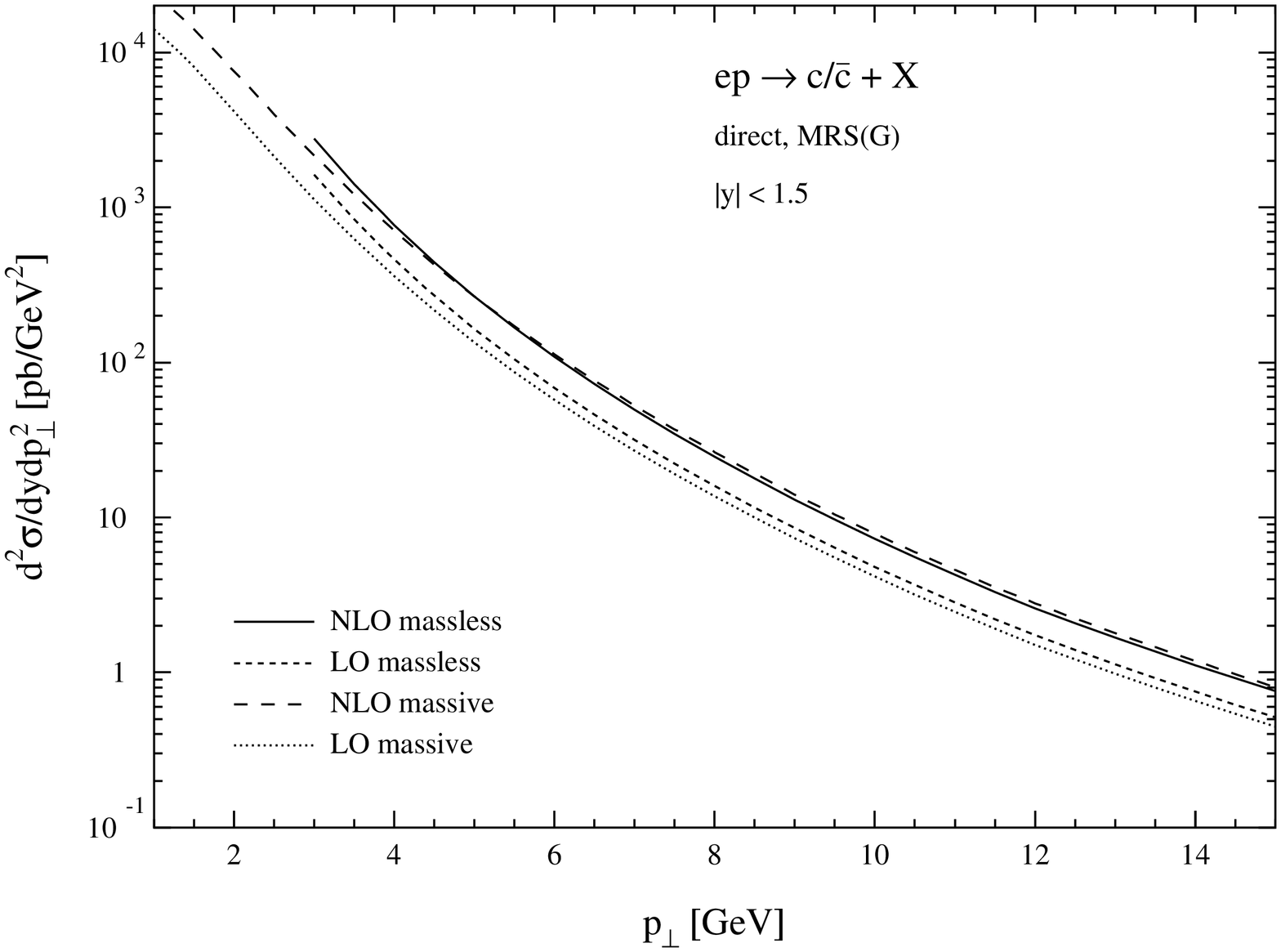,%
height=10.0cm,%
width=14.0cm,%
bbllx=1.0cm,%
bblly=1.9cm,%
bburx=19.4cm,%
bbury=26.7cm,%
rheight=8.2cm,%
rwidth=15cm,%
angle=-90}

\vspace{1.8cm}

 {\bf Fig.1a:}{\em \,Transverse momentum distributions of $ep\to c/\bar{c}+X$
                     averaged over the rapidity range $|y|<1.5$ for the
                     direct contribution in the massless and massive
                     schemes.}

\end{figure}
\begin{figure}[htbp]

\vspace*{-0.5cm}
\hspace*{0.75cm}
\epsfig{%
file=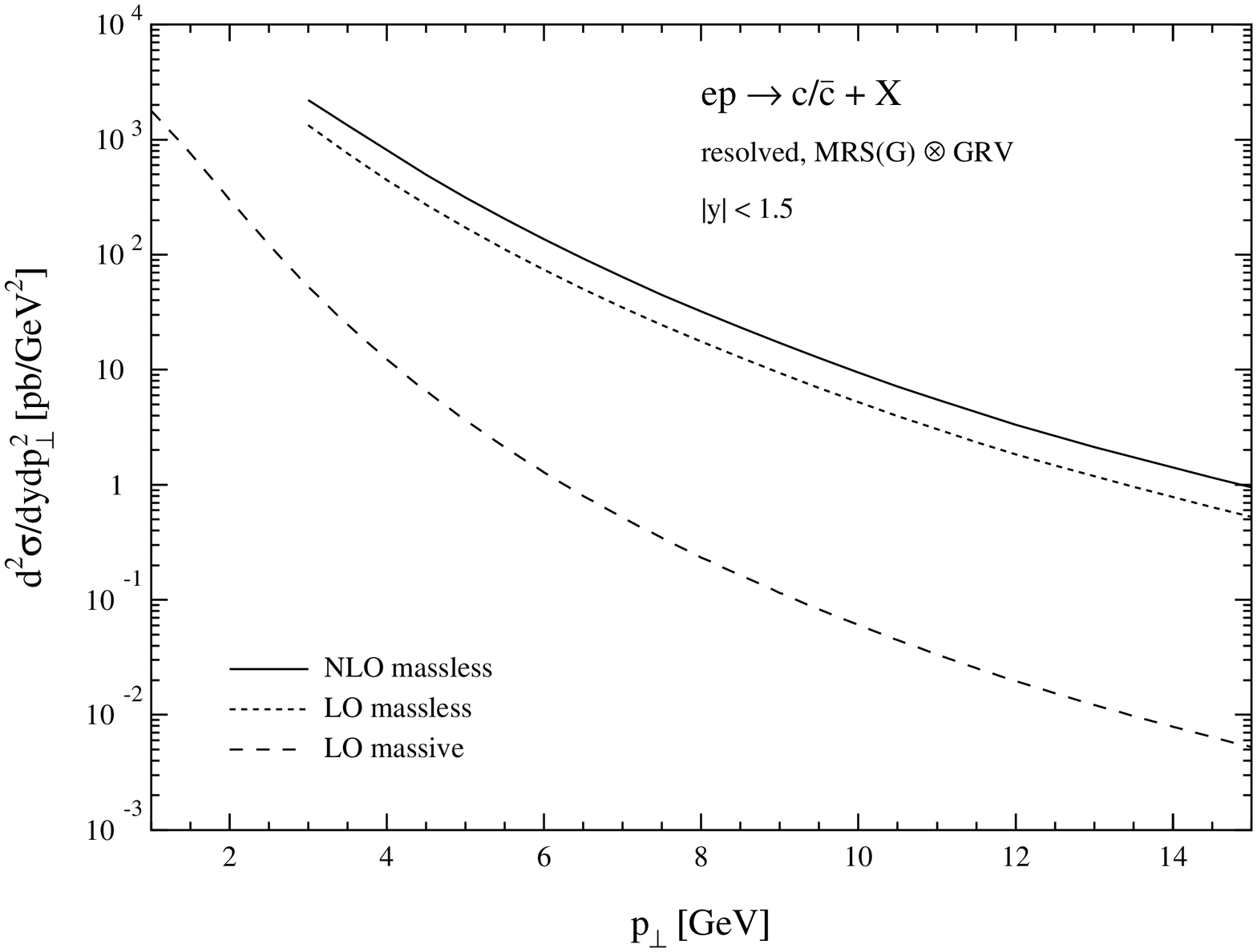,%
height=10.0cm,%
width=14.0cm,%
bbllx=1.0cm,%
bblly=1.9cm,%
bburx=19.4cm,%
bbury=26.7cm,%
rheight=8.2cm,%
rwidth=15cm,%
angle=-90}

\vspace{1.8cm}

 {\bf Fig.1b:}{\em \,Transverse momentum distributions of $ep\to c/\bar{c}+X$
                     averaged over the rapidity range $|y|<1.5$ for the
                     resolved contribution in the massless and massive
                     schemes.}

\end{figure}
\begin{figure}[htbp]

\vspace*{-0.5cm}
\hspace*{0.75cm}
\epsfig{%
file=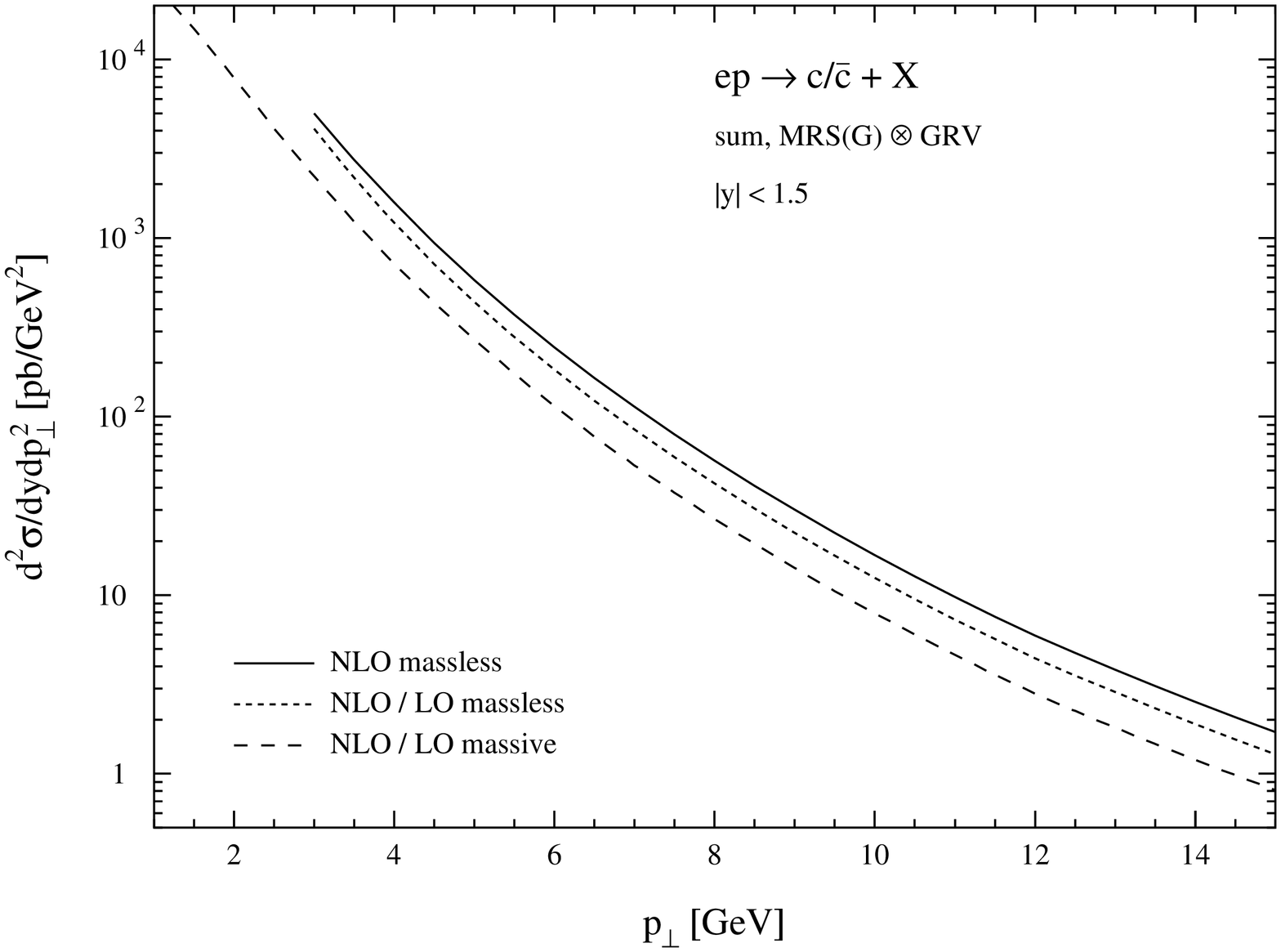,%
height=10.0cm,%
width=14.0cm,%
bbllx=1.0cm,%
bblly=1.9cm,%
bburx=19.4cm,%
bbury=26.7cm,%
rheight=8.2cm,%
rwidth=15cm,%
angle=-90}

\vspace{1.8cm}

 {\bf Fig.1c:}{\em \,Transverse momentum distributions of $ep\to c/\bar{c}+X$
                   averaged over the rapidity range $|y|<1.5$ in the massless
                   and massive schemes. Shown is the sum of the NLO massive
                   direct and the LO massive resolved contributions (NLO/LO
                   massive), the sum of the NLO massless direct and the LO
                   massless resolved contributions (NLO/LO massless), and the
                   sum of the NLO massless direct and the NLO massless
                   resolved contributions (NLO massless).}

\end{figure}

\newpage

\begin{figure}[htbp]

\vspace*{-0.5cm}
\hspace*{0.75cm}
\epsfig{%
file=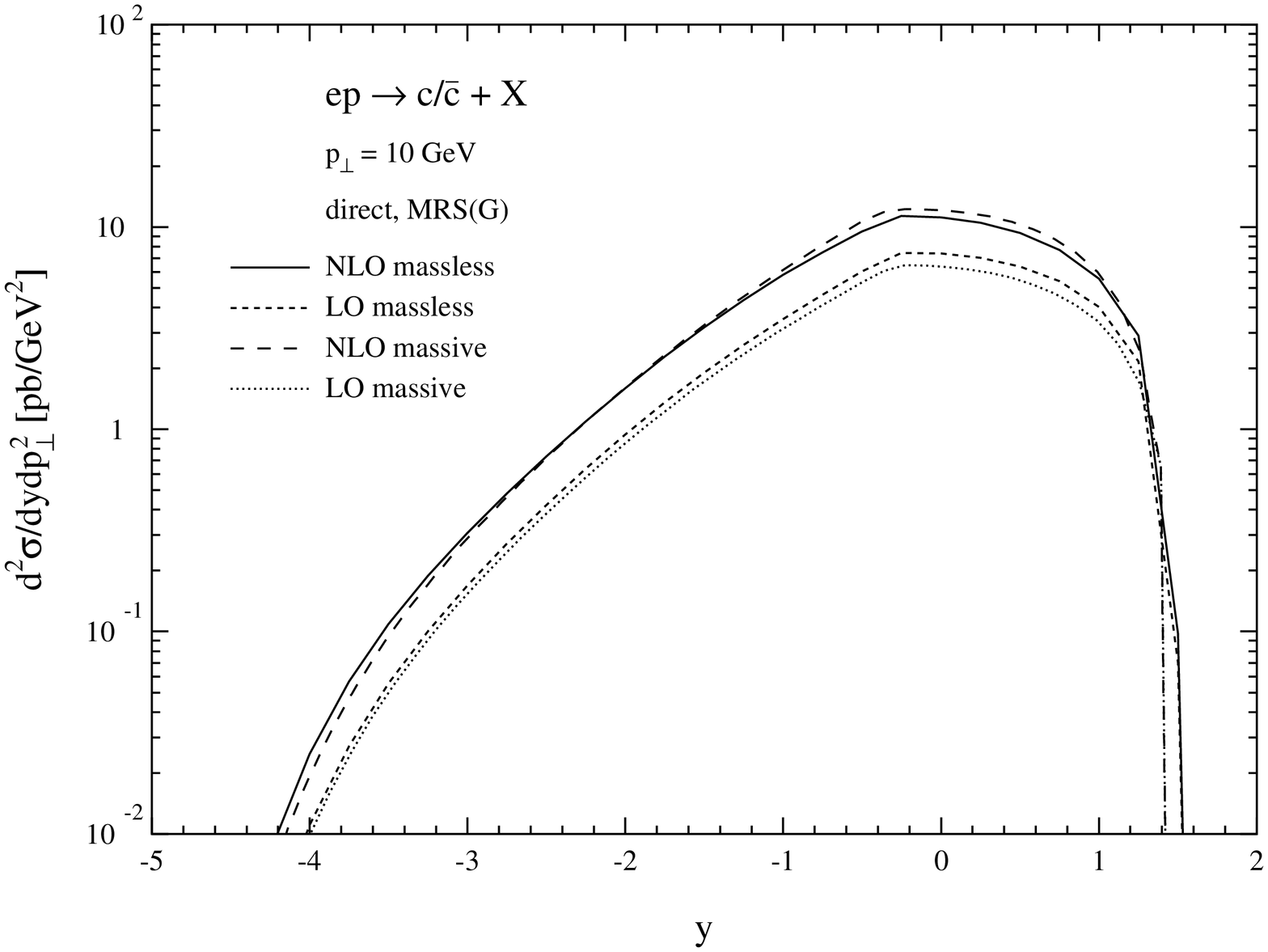,%
height=10.0cm,%
width=14.0cm,%
bbllx=1.0cm,%
bblly=1.9cm,%
bburx=19.4cm,%
bbury=26.7cm,%
rheight=8.2cm,%
rwidth=15cm,%
angle=-90}

\vspace{1.5cm}

 {\bf Fig.2a:}{\em \, Rapidity distributions of $ep\to c/\bar{c}+X$ at
                      $p_\perp = 10 $~GeV for the direct contribution
                      in the massless and massive schemes.}

\end{figure}
\begin{figure}[htbp]

\vspace*{-0.5cm}
\hspace*{0.75cm}
\epsfig{%
file=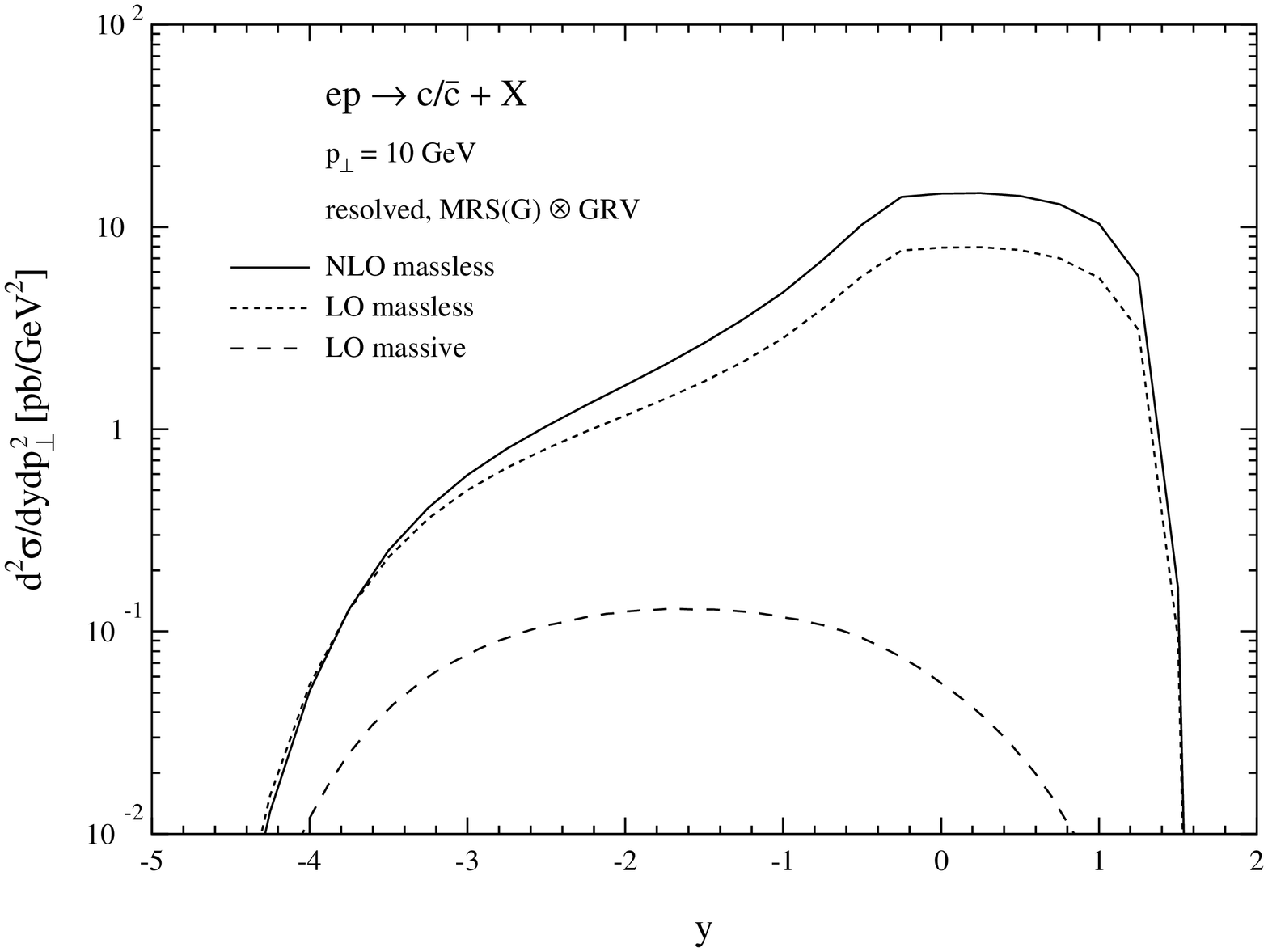,%
height=10.0cm,%
width=14.0cm,%
bbllx=1.0cm,%
bblly=1.9cm,%
bburx=19.4cm,%
bbury=26.7cm,%
rheight=8.2cm,%
rwidth=15cm,%
angle=-90}

\vspace{1.5cm}

 {\bf Fig.2b:}{\em \, Rapidity distributions of $ep\to c/\bar{c}+X$ at
                      $p_\perp = 10 $~GeV for the resolved contribution
                      in the massless and massive schemes.}

\end{figure}
\begin{figure}[htbp]

\vspace*{-0.5cm}
\hspace*{0.75cm}
\epsfig{%
file=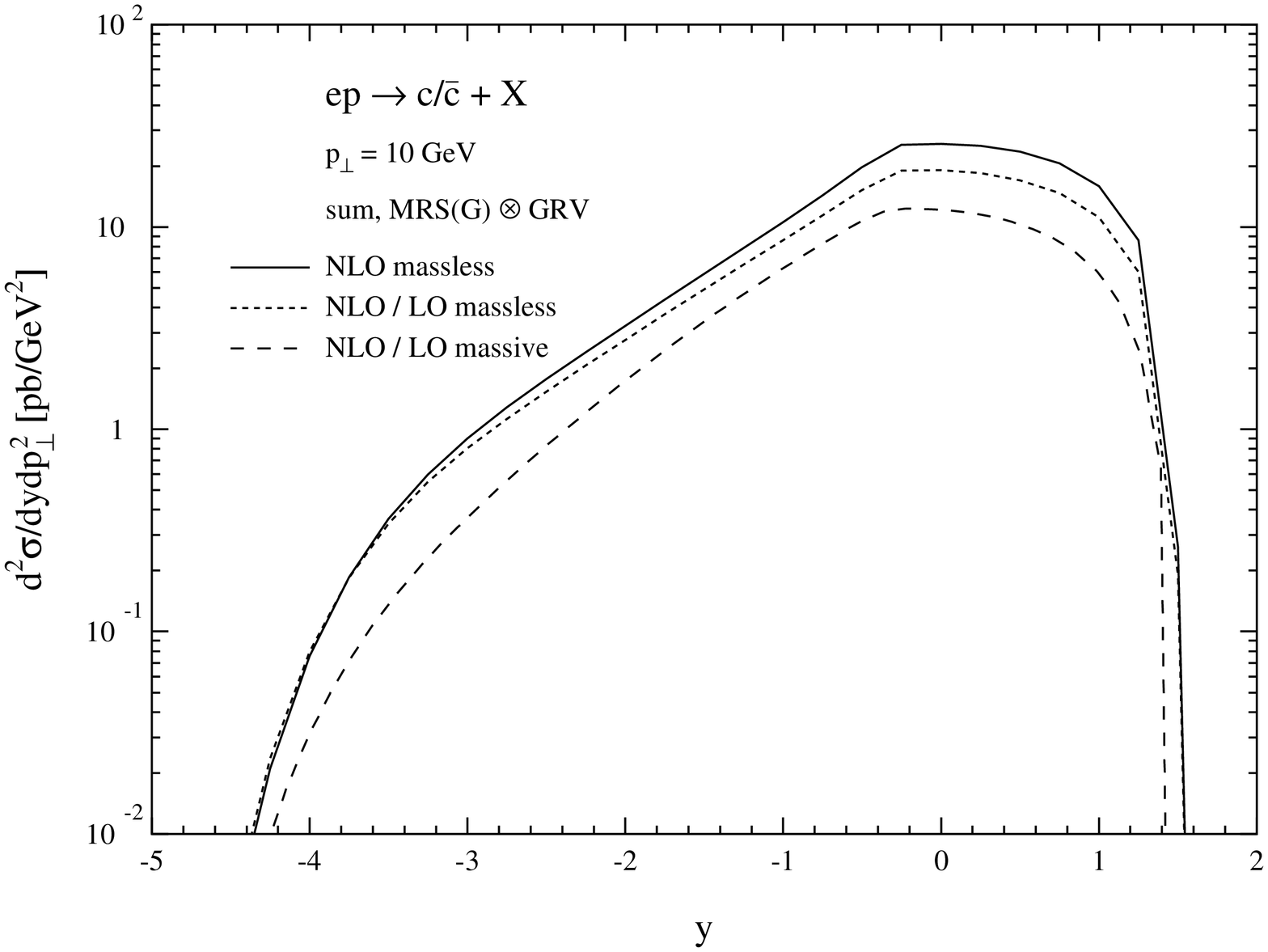,%
height=10.0cm,%
width=14.0cm,%
bbllx=1.0cm,%
bblly=1.9cm,%
bburx=19.4cm,%
bbury=26.7cm,%
rheight=8.2cm,%
rwidth=15cm,%
angle=-90}

\vspace{1.5cm}

 {\bf Fig.2c:}{\em \, Rapidity distributions of $ep\to c/\bar{c}+X$ at
                      $p_\perp = 10 $~GeV for the resolved contribution
                      in the massless and massive schemes. Shown is the
                      sum of the NLO massive direct and the LO massive
                      resolved contributions (NLO/LO massive), the sum of
                      the NLO massless direct and the LO massless resolved
                      contributions (NLO/LO massless), and the sum of the
                      NLO massless direct and the NLO massless resolved
                      contributions (NLO massless).}

\end{figure}

\end{document}